\documentclass[aip,pof,amsmath,amssymb,preprint,]{revtex4-1}
\usepackage{epsfig}
\usepackage{color}
\usepackage{graphicx}
\usepackage{dcolumn}
\usepackage{bm}

\begin{document}

\title{Corrugated interfaces in multiphase core-annular flow}
\author{Ho Cheung Shum}
\affiliation{School of Engineering and Applied Sciences, Harvard University, Cambridge, Massachusetts 02138, USA}

\author{Alban Sauret}
\affiliation{School of Engineering and Applied Sciences, Harvard University, Cambridge, Massachusetts 02138, USA}
\affiliation{\'Ecole Normale Sup\'erieure de Lyon, Lyon 69364, France}

\author{Alberto Fernandez-Nieves}\altaffiliation[Present adress: ]{School of Physics, Georgia Institute of Technology, Atlanta, Georgia 30332, USA.}
\affiliation{School of Engineering and Applied Sciences, Harvard University, Cambridge, Massachusetts 02138, USA}

\author{Howard A. Stone}\altaffiliation[Present adress: ]{Department of Mechanical and Aerospace Engineering, Princeton University, Princeton, New Jersey 08544, USA.}

\affiliation{School of Engineering and Applied Sciences, Harvard University, Cambridge, Massachusetts 02138, USA}

\author{David A. Weitz}\email{weitz@seas.harvard.edu}
\affiliation{School of Engineering and Applied Sciences, Harvard University, Cambridge, Massachusetts 02138, USA}
\affiliation{Department of Physics and Kavli Institute for Bionano Science and Technology, Harvard University, Cambridge, Massachusetts 02138, USA}

\date{20 April 2010}
\begin{abstract}
Microfluidic devices can be used to produce highly controlled and monodisperse double or multiple emulsions. The presence of inner drops inside a jet of the middle phase introduces deformations in the jet, which leads to breakup into monodisperse double emulsions. However, the ability to generate double emulsions can be compromised when the interfacial tension between the middle and outer phases is low, leading to flow with high capillary and Weber numbers. In this case, the interface between the fluids is initially deformed by the inner drops but the jet does not break into drops. Instead, the jet becomes highly corrugated, which prevents formation of controlled double emulsions. We show using numerical calculations that the corrugations are caused by the inner drops perturbing the interface and the perturbations are then advected by the flow into complex shapes.
\end{abstract}

\maketitle

\label{firstpage}

\section{Introduction}

Multiphase flows are widely used industrially for a plethora of applications including oil extraction, formulation of food, and home and personal care products. A good understanding of multiphase flows is needed to facilitate the use of these highly complex systems in the various applications. For example, the use of core-annular flows facilitates water-lubricated transport of heavy viscous oils by providing new approaches to mitigate problems such as fouling of oil pipes.\cite{Joseph1997} Depending on the properties of the flow, such as pressure gradients and fluid velocities, a variety of flow structures including corrugated interfaces has been observed. While the study of jets and drops in multiphase systems has facilitated numerous conventional applications,\cite{Eggers2008} the emergence of microfluidics creates new opportunities to study multiphase flow in microscale systems.\cite{Gunther2006,Guillot2007,Humphry2009}

Microfluidic methods for manipulating multiphase, water-and-oil systems have a great potential for applications involving encapsulation. One particularly promising way to accomplish such encapsulation is through the controlled generation of multiple emulsions,\cite{Chu2007,Chang2008,Okushima2004,Seo2007,Utada2005} which can be done in various confined microfluidic channels with uniform depths, such as T-junctions\cite{Nisisako2005} and flow-focusing channels.\cite{Seo2007,Wan2008} Another approach to form these emulsion is to direct droplet breakup in capillaries aligned axisymmetrically.\cite{Utada2005} The breakup of the inner phase and the middle phase can either happen simultaneously or in two steps. In the latter case, the inner drops are first generated and trigger the breakup of the middle jet into double emulsions due to interfacial tension effects. Tapered capillaries are typically used in these devices so that droplets slow down after generation, resulting in lower viscous stresses, which help prevent destabilization of the droplets due to these stresses. The microfluidic technique allows the formation of monodisperse double emulsion with a high degree of control over the size and shape of the emulsion drops, and with excellent encapsulation efficiency.\cite{Shah2008,Utada2007a} Thus it provides a versatile platform for studying the dynamics of drop formation\cite{Utada2008,Utada2007} and for fabricating novel functional materials.\cite{Shah2008,Shum2008,Shum2008b,Shum2010,Shum2009}

Despite the high degree of control afforded by the technique, the multiple emulsions formed are typically not stable due to coalescence between miscible phases. For example, in a double emulsion, the innermost drop can coalesce with the outermost, continuous-phase fluid, destroying the multiple emulsion. A common strategy to improve emulsion stability is to introduce surfactants, which generally help to suppress coalescence. However, the introduction of surfactants also leads to lower interfacial tensions; the dynamics associated with these lower interfacial tensions can result in a variety of fluid flow patterns and instabilities that disrupt the regular formation of the double emulsion. Consequently, these effects can severely affect the encapsulation efficiency and prevent prolonged operation of the processes, which makes microfluidic emulsification unfeasible for industrial applications. It is, therefore, important to understand the origin and mechanisms of formation of these flow patterns.

In this paper, we study the corrugations observed along the fluid-fluid interface inside a glass capillary device when the interfacial tension between the middle phase and the outer phase is very low, due to the introduction of surfactants. We operate the device in a regime where the innermost fluid forms drops by dripping right at the orifice, whereas the middle fluid forms a jet. We study experimentally the interfacial corrugations, which are initially triggered by advection of the innermost drops inside the jet. Due to the low interfacial tension between the middle and outer phases, the perturbations of the interface can be advected into complex shapes by the flow. We observe the phenomenon when the surface tension is low compared to both inertial and viscous forces; as a result, both the capillary and Weber numbers are very large. We suggest an explanation of the mechanism of formation and evolution of the corrugations, which is validated by reproducing the corrugations in a numerical model that agrees qualitatively and quantitatively with the experiments.


\section{Experimental}

\begin{center}
\begin{figure}
\begin{center}\includegraphics[width=12cm]{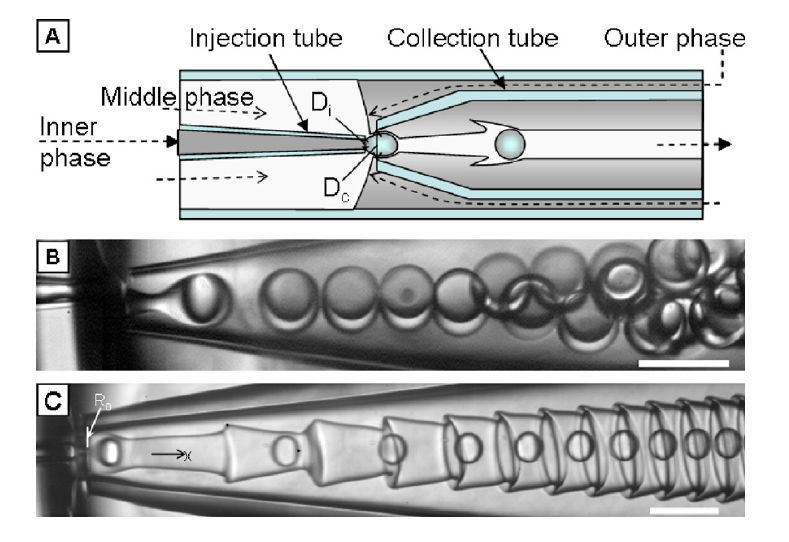}\end{center}
\caption{Double emulsions and the formation of corrugated interfaces. (a) Schematic of the experimental setup. (b) Generation of monodisperse double emulsion droplets. The inner, middle, and outer fluids are, respectively, water, 10$\%$ (w/w) Span 80 in dodecane, and water at flow rates of $1000$, $1500$, and $5000$ $\mu$L h$^{-1}$, respectively. In this system, the interfacial tension between the middle and outer phases is $\gamma=2.7 \pm 0.2$ mN m$^{-1}$, $Re \simeq 15$, $Ca \simeq 0.15$, and $W\!e \simeq 0.6$. (c) Formation of corrugations when the interfacial tension between the middle and outer phases is significantly reduced. The inner, middle, and outer fluids are water, $2 \%$ (w/w) Span 80 in dodecane and water with $4$ mM SDS and the flow rates are $500$, $10^4$, and $10^4$	$\mu$L h$^{-1}$ for the inner, middle, and outer fluids, respectively. In this system, the interfacial	 tension between	the middle	and	outer phases	is	30 $\pm$ 4	$\mu$N m$^{-1}$, $Re \simeq 38$, $Ca \simeq 37$, and $W\!e \simeq 600$. The radius of the capillary at its entrance $R_0$ is $38.8$ $\mu$m. Scale bars are $100$ $\mu$m.}
\label{figure1}
\end{figure}
\end{center}

All experiments were carried out in glass microcapillary
devices,\cite{Utada2005} shown schematically in Fig. \ref{figure1}(a). The circular capillaries, with inner and outer diameters of $0.58$ and $1.0$ mm (World Precision Instruments, Inc.), were tapered to the desired shapes with a micropipette puller (P-97, Sutter Instrument, Inc.) and a microforge (Narishige International USA, Inc.). Two tapered capillaries with different tip diameters were aligned inside a square glass capillary (Altantic International Technology, Inc.) with an inner dimension of $1.05$ mm. The innermost dispersed phase flows through the round capillary with the smaller-diameter tip; this is the injection tube. The second round capillary, with the larger-diameter tip collects the double emulsion and directs it to the exit of the device. The diameter $D_c$ of the collection tube ranged from $80$ to $200$ $\mu$m, while the diameter $D_i$ of the injection tube ranged from $40$ to $100$ $\mu$m. A transparent epoxy resin (5 Minute Epoxy, Devcon) was used to seal the capillaries where necessary. Positive syringe pumps (PHD 2000 series, Harvard Apparatus) were used to deliver the different phases at the desired flow rates. The flow behavior inside the devices was monitored with a 10$\times$ objective using an inverted microscope (Leica, DMIRBE) equipped with a high-speed camera (Phantom, V5, V7, or V9). Unless otherwise noted, the inner, middle, and outer phases consisted of water, $2\% - 10\%$ (w/w) Span 80 surfactant in dodecane, and water with $0 - 8$ mM sodium dodecyl sulfate (SDS), respectively. Solutions and solvents were all filtered before introduction into the glass microcapillary devices.

Unless otherwise stated, all chemicals were purchased from Sigma-Aldrich. Water with a resistivity of 18.2 M$\Omega$ cm$^{-1}$ was acquired from a Millipore Milli-Q system. Whenever possible, characteristic interfacial tensions were measured by forming a pendant drop of the denser phase at the tip of a blunt stainless steel needle (McMaster-Carr, 20 Gauge) immersed in the second phase and performing drop-shape analysis.\cite{DelRio1997} Low interfacial tensions that could not be precisely determined using the pendant drop method were measured using a spinning drop tensiometer (SITE 100, Kruss GmbH). Viscosities were measured with a Ubbelohde capillary viscometer mounted in a temperature-controlled water bath.

\section{Results and discussion}

In a typical system for generation of emulsions, the inner phase is typically at a low flow rate, allowing droplets of the
inner phase to be formed at the orifice of the injection tube in a flow regime known as dripping. By contrast, the middle
phase is typically injected at high flow rates; as a result, a jet of the middle phase is formed inside the collection tube; thus it is operated in a regime known as jetting. The advection of the inner droplets inside the middle jet causes the interface between the middle and the outer phases to deform. With a typical interfacial tension, for example, $\gamma=2.7 \pm 0.2$ mN m$^{-1}$, for the experiment shown in Fig. \ref{figure1}(b), the deformation grows
and eventually causes the middle jet to break up, thereby forming double emulsion drops. The breakup of the middle
jet occurs as a consequence of the interfacial tension and viscous stresses due to the flow of the continuous phase.
While interfacial tension tends to favor droplet formation, viscous stresses induce the interface to follow the flow profile inside the channel. The balance of these two effects is characterized by the capillary number of the continuous
phase:	$Ca = \eta_{out}\,V_{out}/\gamma$,	where	$\eta_{out}$ and $V_{out}$ are the viscosity and velocity of the outer continuous phase, respectively. The flow in the experiment shown in Fig. \ref{figure1}(b) has a low capillary
number,	$Ca \simeq 0.15$. Apart from viscous stresses, inertial effects can also compete with surface tension to prevent droplet breakup. If the jet moves faster than the continuous phase, the velocity difference induces a large shear at the interface that decelerates the jet, causing it to widen. The balance of this inertial effect and surface tension is characterized by the Weber number of the jet-forming phase, which includes both the inner and middle phases: $W\!e=\rho_{in} \,d\,{V_{in}}^2/\gamma$, where $\rho_{in}$ and $V_{in}$ are the density and velocity of the jet-forming phase, respectively. The flow in the experiment shown in Fig. \ref{figure1}(b) has a low Weber number,	$W\!e \simeq 0.6$. 

However, the performance of the glass microcapillary devices in generating controlled double emulsions is compromised when the interfacial tension between the middle and the outer phases is low such that $Ca$, $W\!e \geq O(1)$. Under such flow conditions, the middle jet does not break up; instead a highly corrugated interface is formed, as shown in Fig. \ref{figure1}(c). Here the interfacial tension is $30 \pm 4$ $\mu$N m$^{-1}$, $Ca \simeq 37$, and $W\!e \simeq 600$; both $Ca$ and $W\!e$ are about $100$ times the values for the flow shown in Fig. \ref{figure1}(b).

\begin{center}
\begin{figure}
\begin{center}\includegraphics[width=10cm]{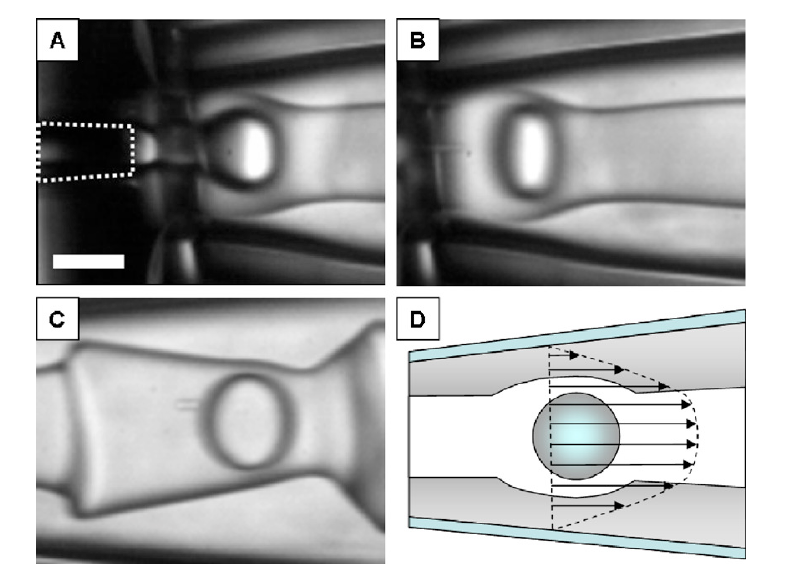}\end{center}
\caption{(a) Formation of an inner droplet in the dripping regime. The dotted white line indicates the location of the injection tube. The scale bar is $100$ $\mu$m and it applies to [(a)-(c)]. (b) Deformation of the middle jet by the pinch-off of the inner droplet. (c) Evolution of the deformed middle jet into a corrugation. (d) Proposed mechanism of the corrugation formation. The interface between the middle and outer phases is initially deformed by the droplet and then evolves following the parabolic flow configuration due to the flow in a widening channel.}
\label{figure2}
\end{figure}
\end{center}

To understand the formation of the highly corrugated interface, we examine the onset of the corrugation with optical microscopy. Since the interfacial tension between the innermost droplet phase and the middle phase is about two orders of magnitude higher than that between the middle phase and the outermost continuous phase, the interface between the middle and outermost phases is significantly more deformable than that between the innermost and middle phases. As a result, when an inner droplet is about to pinch off from the injection tube in the dripping regime, the interface between the middle and the outer phases starts to become deformed. At this point, the inner droplet is still attached to the tip of the injection tube; however, the droplet size is already larger than the size of an unperturbed middle jet. This geometry causes the middle jet to curve around the inner droplet, as shown in Fig. \ref{figure2}(a).

\begin{center}
\begin{figure}
\begin{center}\includegraphics[width=11cm]{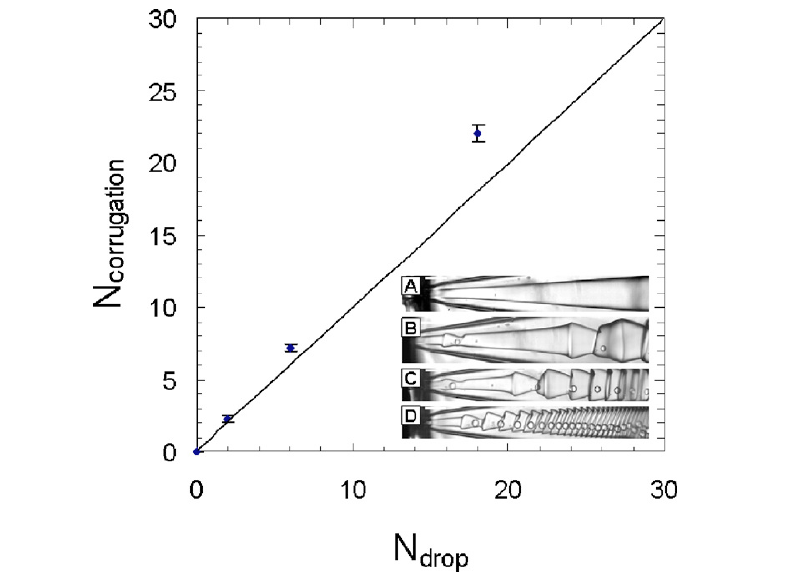}\end{center}
\caption{Number of corrugations observed, $N_{corrugation}$, as a function of the number of inner drops, $N_{drop}$, for a given length of the device. The solid line has a slope of unity. Since the inner drops flow at a higher speed than the corrugations and the overall speed decreases as they flow downstream in the widening channel, the number of corrugations counted in a typical image is a little larger than that of the inner drops. Each data point was obtained by counting the number of corrugations and the number of inner drops in at least five images. Inset: Snapshots of the flow when the flow rates of the inner phase are (a) 0, (b) 5, (c) 50, and (d) 500 $\mu$L h$^{-1}$ while keeping the flow rates of the middle and outer phases at $10^4$ $\mu$L h$^{-1}$ and	$1.5\times10^4$ $\mu$L h$^{-1}$, respectively.	Scale bars are 100 $\mu$m.}
\label{figure3}
\end{figure}
\end{center}

After the complete pinch-off, the inner droplet continues to flow downstream at a higher velocity than the deformed interface, which lags behind, as shown in Fig. \ref{figure2}(b). As the fluids flow downstream, the shape of the deformed interface evolves such that the front of the deformation has a smaller diameter than its back [Fig. \ref{figure2}(c)]. We also measure the number of corrugations as a function of the number of inner drops generated, as shown in Fig. \ref{figure3}. The frequency of the corrugations approximately matches the generation frequency of the inner droplets. These observations suggest that the corrugations are initiated by the formation of the inner drops, rather than simply arising as an intrinsic fluid or interfacial instability such as the Rayleigh capillary instability. While the formation of the corrugations is triggered by the motion of the inner drops, the high curvature in the resultant highly corrugated jet is sustained because of the ultralow interfacial tension of the interface.

Two potential driving forces for the evolution of the interface to its final shape are the disturbance flow due to the moving inner drop and advection of the initially deformed interface in a pressure-driven Poiseuille flow in a widening capillary. If the former effect dominates, the inner droplet should continue to significantly deform the interface between the middle and outer fluids after the drop formation. However, the magnitude of the deformation caused directly by the drop is inconsequential when compared to that of the corrugation, as suggested by the largely undeformed interface near the inner droplet in Fig. \ref{figure2}(c). Thus, it seems unlikely that the disturbance flow due to the moving inner droplet is the main driving force. Therefore, we hypothesize that the final shape of the corrugations is caused by advection of an initially perturbed interface in a Poiseuille flow in a widening channel, as illustrated schematically in Fig. \ref{figure2}(d).

Based on these observations, we suggest a mechanism for the formation and evolution of the corrugations: the inner fluid breaks up into droplets and due to the incompressibility of the middle fluid, the interface between the middle and outer fluids is deformed into a nearly spherical shape around the sphere. Because of the low interfacial tension, the viscous stresses and inertial forces dominate. Thus the deformed interface passively flows downstream according to the two-phase Poiseuille flow profile. Then, due to the approximately parabolic velocity distribution, the part of the curved interface that is closer to the channel wall flows at a lower velocity than the part near axis of the channel. As a result, the deformed interface develops into arrowlike, folded corrugations, as shown in Fig. \ref{figure1}(c).

\begin{center}
\begin{figure}
\begin{center}\includegraphics[width=10cm]{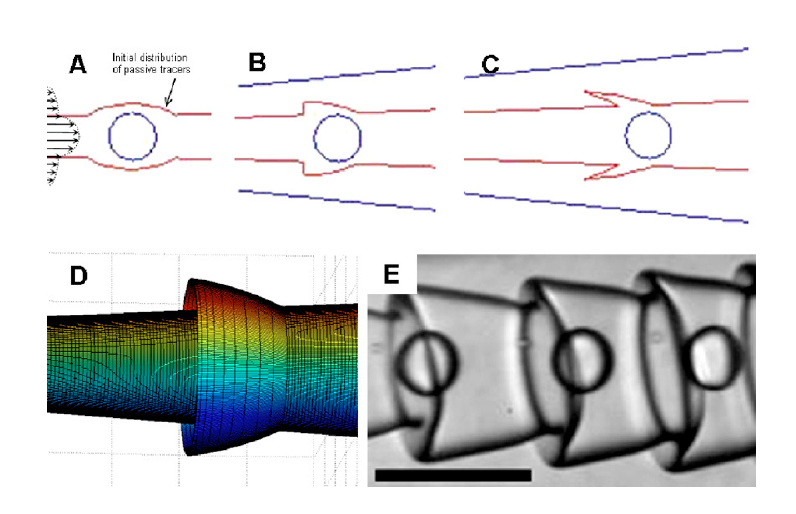}\end{center}
\caption{(a) The initial shape of the interface after pinch-off of an inner droplet. The shape is used as the initial condition for the numerical calculation. (b) Shape of the interface developed under pressure-driven flow in the widening channel. (c) Shape of the interface further downstream from (b). (d) Three-dimensional shape of the interface by rotating the two-dimensional shape in (c) by $360 \!\char23$. (e) Image of the corrugations observed experimentally. The shape of the corrugations is captured qualitatively by the calculation, as shown in (d). Scale bar is 100 $\mu$m.}
\label{figure4}
\end{figure}
\end{center}

To confirm this proposed mechanism, we construct a model for the occurrence of the interfacial corrugations of the experimental system that highlights the most important ideas. In our experimental system, the viscosities of the middle and the outer phases are $1.44$ and $1.0$ cP, respectively. Since the viscosity ratio is close to one, we assume a single-phase pressure-driven velocity field in our numerical calculations; this simplifies the calculations significantly while still capturing the main features of the experimental observations. The initial condition consists of two lines of imaginary tracer particles that approximate the initial shape of the experimentally observed interface immediately after pinch-off of the inner droplet, as shown in Figs. \ref{figure2}(a) and \ref{figure4}(a). The only role that the inner drop plays in our calculations is to initially perturb the interface. The tracers are then advected by the parabolic velocity distribution in the widening channel
\begin{equation}
U(r)=U_{max}\,\Biggl[1-\left(\frac{r}{R(z)}\right)^2\Biggr]
\end{equation}
where	$R(z)=\alpha\,z+R_0$ is the inner radius of	the capillary at	the axial distance $z$ from the entrance of the capillary and $r$ is the radial distance of the tracer from the centerline. The capillary has a radius of $R_0$ at its entrance, and widens with a slope of $\alpha$, which has a typical value of about $0.23$, as measured experimentally.

As the perturbed interface flows downstream, the tracers that are furthest from the capillary axis lag behind those that are closer to the capillary axis. As a result, the initially spherical deformation sharpens, as shown in Fig. \ref{figure4}(b). The part of the interface further away from the axis continues to lag behind and the corrugation folds, as shown in Fig. \ref{figure4}(c). The three-dimensional representation of the shape based on the results from our numerical calculations closely resembles the experimentally observed shape, as shown in Figs. \ref{figure2}(d) and \ref{figure4}(e).

\begin{center}
\begin{figure}
\begin{center}\includegraphics[width=12cm]{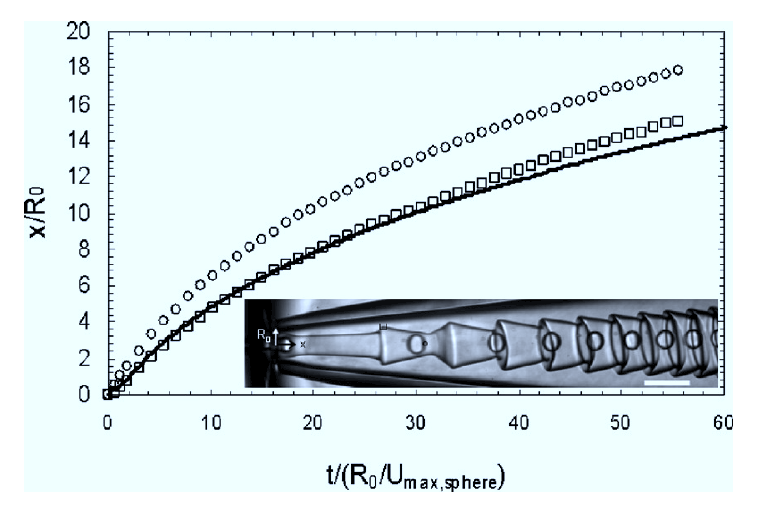}\end{center}
\caption{Measured positions of the inner drop ($\bigcirc$) and the tip of the corrugation ($\Box$) as	a	function	of	normalized	time, $t / (R_0 / U_{max,sphere})$. The solid line is the prediction for the position of the tip of the corrugation from the numerical calculation with no fitting parameters. Inset: a snapshot of the corrugated interface, with the positions of the inner drop ($\bigcirc$) and the tip of the corrugation ($\Box$) marked. The radius of the capillary widens from $R_0$ at the entrance at a slope of about $0.23$. Scale bar is 100 $\mu$m.}
\label{figure5}
\end{figure}
\end{center}

We also track the positions of the tip of a corrugation obtained from the numerical calculations over time, neglecting the effects of interfacial tension. For a comparison with the experimental results, we normalize both the experimentally obtained and numerically calculated data. The $x$-positions of the tip are normalized by the radius of the capillary at the entrance, $R_0$. The timescales are normalized by a reference timescale, ($R_0 / U_{max,sphere}$), where $U_{max,sphere}$ is the speed of the sphere at the entrance of the channel. We then plot the normalized positions of an inner drop and the tip of a corrugation as a function of normalized time in Fig. \ref{figure5}. The experimental observed positions of the tip, represented by the square symbols, agree very well with the numerically calculated positions of the tip, shown with a solid line in the plot. This result supports our hypothesis that the mechanism that produces the corrugations is folding of an initially perturbed shape by the slowly varying parabolic velocity profile. The result of the calculations deviates slightly from experiments at long time, as shown by the increasing gap between the symbols and the solid line in Fig. \ref{figure5}. The deviation likely results from effects on the flow due of the motion of the inner droplets, which have not been included in our simple model.

\begin{center}
\begin{figure}
\begin{center}\includegraphics{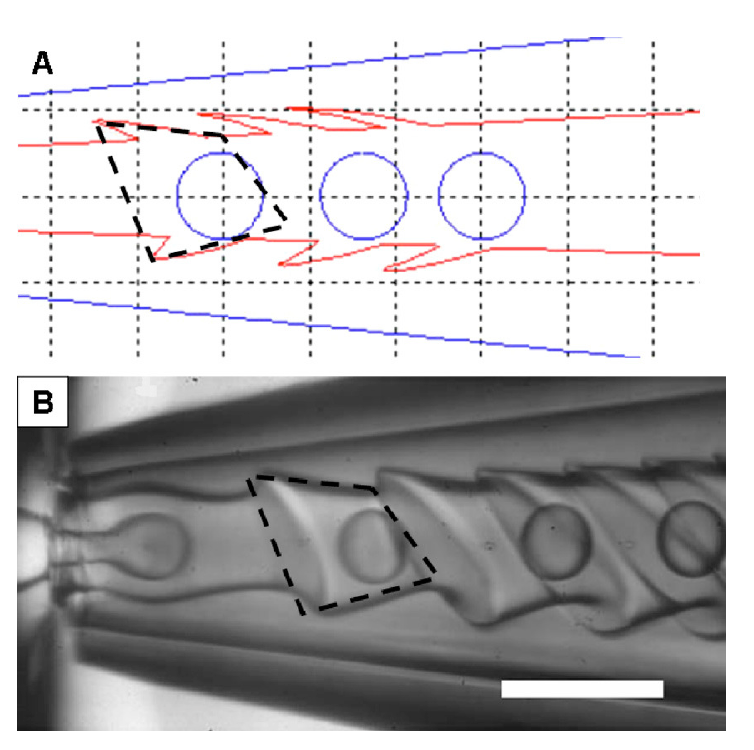}\end{center}
\caption{(a) Numerical calculations of the corrugations on the surface of a misaligned middle jet. (b) Experimentally observed corrugations on the surface of a slightly off-centered middle jet. Misalignments of the capillaries in the device lead to axial asymmetry in the positions of the different phases, often resulting in asymmetric corrugations. Such asymmetry is captured by the calculation. To highlight the close resemblance, a box of the same size and shape drawn with black dotted lines are placed around both the experimentally observed and the numerically calculated corrugations in A and B. Scale bar is $100$ $\mu$m.}
\label{figure6}
\end{figure}
\end{center}

To test the robustness of the proposed mechanism, we also numerically calculate the shape of the perturbed interface obtained in a device where the injection and collection tubes are slightly off-center. This feature is often observed experimentally and leads to axial asymmetry in the droplet generation process. As a result of the misalignment, the inner drops are closer to one side of the collection tube and the initial shape of the interface is changed accordingly. To confirm the validity of our model, we perform our numerical calculations with an initial condition where the two lines of imaginary tracer particles are asymmetrically positioned to approximate the initial shape of the interface deformed by an off-center inner drop. From the results of our calculation, we obtain corrugations that are axially asymmetric about the capillary axis, as shown in Fig. \ref{figure6}(a). The shape obtained is similar to the experimentally observed asymmetric corrugations shown in Fig. \ref{figure6}(b). To highlight the close resemblance, the boxes of the same size and shape, drawn with black dotted lines, are placed around both the experimentally observed and the numerically calculated corrugations in Figs. \ref{figure6}(a) and \ref{figure6}(b).

To summarize, we have observed corrugated interfaces in a multiphase core-annular flow at low interfacial tension in glass capillary microfluidic devices designed for the generation of double emulsions. We present qualitative evidence that the initial perturbations of the compound jet are introduced by the advection of the small inner drops inside the jet. As a result of the low interfacial tension between the middle and outer phases, the perturbations are advected into complex shapes by the flow; the high curvature of the resultant highly corrugated shape is not suppressed since the Weber numbers are large (low values of the surface tension). We also perform numerical calculations in which a single-phase pressure-driven velocity field advects tracers that outline the interface. Despite the simplifying assumption, our calculations capture the qualitative and quantitative behaviors of the experimentally observed corrugations. The improved understanding of this phenomenon will help provide a higher degree of control over multiple emulsions generated with microfluidic techniques. Moreover, our study also suggests a new route to create highly corrugated interfaces. With an appropriate strategy for solidifying the middle jet, our system may lead to fibers with significantly enhanced surface area for biomedical and catalytic applications that requires large surfaces for improved activities.

\section*{Acknowledgements}
This work was supported by the NSF (Contract No. DMR-1006546) and the Harvard MRSEC (Contract No. DMR-0820484). H.C.S. and D.A.W. were supported in part by BASF Ludwigshafen in Germany. A.F-N. thanks Ministerio de Educacion y Ciencia (Contract No. DPI2008-06624- C03-03) and University of Almeria. We thank Andy Utada and Daeyeon Lee for helpful discussions and Kruss GmbH for the spinning drop tensiometer measurements.

\bibliography{biblio}

\end{document}